\documentclass[12pt, a4paper, twoside]{fec}

\usepackage{amsmath}
\usepackage{amsfonts}
\usepackage{amsthm}
\usepackage{graphicx}
\usepackage{amssymb}
\usepackage{color}

\usepackage{wrapfig} 

\newcommand{\grad}{\nabla}
\newcommand{\Perp}{{\mbox{$\scriptscriptstyle \perp$}}}

\DeclareMathAlphabet{\mathpzc}{OT1}{pzc}{m}{it}

\newcommand{\incfig}{\centering\includegraphics}
\setkeys{Gin}{width=0.9\linewidth,keepaspectratio}

\newcommand{\mvec}[1]{\mathbf{#1}}
\newcommand{\gvec}[1]{\boldsymbol{#1}}

\newcommand{\gke}{{\tt Gkeyll}}

\newcommand{\ignore}[1]{}  

\renewcommand{\baselinestretch}{0.98}

\title{Scrape-Off Layer Turbulence in Tokamaks Simulated with a
  Continuum Gyrokinetic Code}

\author[1]{A.~Hakim}
\author[2]{E.L.~Shi}
\author[3]{I.G.~Abel}
\author[1]{G.W.~Hammett} 
\author[2]{T.~Stoltzfus-Dueck}
\affil[1]{Princeton Plasma Physics Laboratory, Princeton, NJ 08543-0451}
\affil[2]{Princeton University, Princeton, NJ 08544}
\affil[3]{Princeton Center for Theoretical Science, Princeton University, Princeton, NJ 08544}

\doi

\emailID{ahakim@pppl.gov}

\PaperNumber{TH/P6-2} 

\begin{document}
\maketitle

\begin{abstract}
  We are developing a new continuum gyrokinetic code, \gke, for use in
  edge plasma simulations, and here present initial simulations of
  turbulence on open field lines with model sheath boundary conditions.
  The code implements an energy conserving discontinuous Galerkin scheme,
  applicable to a general class of Hamiltonian equations.  Several
  applications to test problems have been done, including a calculation
  of the parallel
  heat-flux on divertor plates resulting from an ELM crash in JET, for a
  1x/1v SOL scenario explored previously, where the ELM is modeled as a
  time-dependent intense upstream source.
  Here we
  present initial simulations of turbulence on open field
  lines in the LAPD linear plasma device.  We have also done simulations
  in a helical open-field-line geometry.  While various simplifications
  have been made at present, this still includes some of the key physics
  of SOL turbulence, such as bad-curvature drive for instabilities and
  rapid parallel losses with sheath boundary conditions.  This is useful
  for demonstrating the overall feasibility of this approach and for
  initial physics studies of SOL turbulence. We developed a novel
  version of DG that uses Maxwellian-weighted basis functions while
  still preserving exact particle and energy conservation. 
  The Maxwellian-weighted DG method achieves the same error with
  4 times less computational cost in 1v, or 16 times lower cost in the 2
  velocity dimensions of gyrokinetics (assuming memory bandwidth is the
  limiting factor).
\end{abstract}

\vspace{-0.5em}
\section{Introduction}

The edge region of a magnetically confined fusion device (from inside
the top of the pedestal outwards, through the separatrix to the open
field-line scrape-off-layer (SOL) and walls) is very important for
understanding H-mode accessibility conditions, the height of the
pedestal and thus the level of fusion performance, the impact of ELMs
and disruptions and methods to control or mitigate them, the width of
the SOL and heat load on divertor plates, and how much performance can
be improved with lithium walls.

While a lot of progress with sophisticated continuum gyrokinetic codes
has been made in
understanding turbulence in the main
core region of tokamaks, these codes are highly optimized for the
core, and new codes or major extensions are needed to handle the
additional complications of the edge region in a numerically stable and
efficient way.
Computational challenges in the edge include the need to
handle large amplitude fluctuations with steep gradients while avoiding
negative overshoots,
magnetic fluctuations near the beta limit,
open and closed field lines and X-points, strong sources and
sinks from atomic physics and plasma-wall interactions, sheath boundary
conditions, a wide range of time and space scales and of collisionality
regimes.
This requires full-F algorithms that do not
assume the plasma is near-Maxwellian.
%
%


We are developing a new continuum code \gke\ for the edge region,
employing various advanced numerical algorithms, including some novel
versions of Discontinuous Galerkin (DG) algorithms, that
can significantly help with the computational challenges of the edge
region. 
DG has been extensively developed and used in the computational fluid
dynamics and applied mathematics community in the past 15
years\cite{Cockburn:2001vr},
as they combine some of the
advantages of finite-element schemes (low phase error, high accuracy,
flexible geometries) with finite-volume schemes (limiters to preserve
positivity/monotonicity, locality of computation for parallelization).
Some of the standard advection algorithms widely used in fusion
research, while they may have good accuracy or conservation properties,
can have negative overshoots in their solutions, which can cause
problems in the edge region in various ways (and may cause problems with
the sheath).  One of the interesting features of the version of DG we
use is that it can conserve energy exactly even when limiters are used
on the fluxes at cell boundaries to ensure the positivity of the
cell-averaged solutions.  There are still some subtle issues involved in
trying to preserve positivity everywhere within a DG cell, which causes
some numerical heating in the present code, but we
are working on ways to reduce this.  But otherwise, the current
algorithm appears to be robust and numerically stable overall, including
in its interaction with the sheath.

There are other projects developing edge
simulations as well (such as the XGC particle-in-cell code\cite{Ku2016},
which is the only gyrokinetic turbulence code at present capable of
handling open
and closed field lines simultaneously), 
but it is essential to have independent codes to
cross-check each other and accelerate progress,
especially for difficult chaotic problems like edge turbulence.
Different algorithms have different properties and tradeoffs regarding
various features of the solution on the long time scale where
the dynamics becomes chaotic and short-time convergence tests alone
are not necessarily sufficient.

Here we will show initial results from \gke\ of
gyrokinetic simulations of turbulence on open field lines with boundary
conditions that model the sheath.  We are using various
simplifications at present, such as a simple helical magnetic field
model of a SOL, but this is sufficient to demonstrate the overall
feasibility of this approach and the ability to handle computational
difficulties that have been challenging for previous attempts.  Our work
builds on pioneering fluid studies of SOL turbulence by
Ricci and Rogers et al.\cite{Rogers2010,Ricci2010} with the GBS code and
by Popovich, Friedman et al.\cite{Popovich2010,Friedman2013} with 
the BOUT++ code.  Indeed, an important step to make this successful was
finding a gyrokinetic generalization of sheath models used in earlier
fluid simulations as a starting point.

The advanced Discontinuous Galerkin (DG) algorithms being developed for
fusion applications here can also be used for a wide range of
kinetic problems.  Versions of \gke\ are being used
to do full Vlasov-Maxwell (not gyrokinetic) studies in
space physics and Vlasov-Poisson studies of plasma-material interactions
in plasma thrusters (for satellites).  The underlying \gke\ framework
is being used for finite-volume multi-moment extended MHD
simulations in the Princeton Center for Heliophysics.

\vspace{-0.5em}
\section{Numerical Methods}
\label{Numerical}

%
%
%




\gke\ uses a novel energy-conserving, mixed discontinuous Galerkin
(DG)/continuous Galerkin (CG) scheme that conserves energy exactly (in the 
continuous time limit or implicit case) for Hamiltonian systems.
A discontinuous Galerkin
representation is used for the particle distribution function $f$ and a
continuous Galerkin representation is used for the fields $\phi$ and the
Hamiltonian $H$.  This scheme is applicable to
the solution of a broad class of kinetic and fluid problems, described
by a Hamiltonian evolution equation $f_t - \{H,f\} = 0$.
The algorithms used are an extension of the mixed
discontinuous/continuous Galerkin
scheme presented in \cite{Liu2000} for the 2D incompressible Euler
equations. The proofs of conservation of quadratic invariants have
been extended to general Hamiltonian systems.
In particular, it can be be shown, that
the \emph{spatial scheme} conserves energy exactly even with
\emph{upwind fluxes} on the cell boundaries. 
%
%

While it is possible to ensure positivity of the cell-averaged
distribution function with limiters on the fluxes at cell
boundaries, the solution inside a cell can still go negative locally.
We have developed a
framework for exponential basis functions that would avoid
this, though it has not yet been implemented in the main code.  In
the mean time, we implemented correction steps to redistribute
$f$ within a cell to ensure positivity locally 
as well.  Simple
application of such a positivity-correction step can cause significant
numerical heating in some cases.  We have added an
energy correction operator to reduce this, and are in the process of
testing some alternatives to improve this further.  This is similar to
the correction steps implemented in \cite{Taitano:2015} and shares the
same philosophy:
they only modify the algorithm at the
level of the truncation error,
so they do not affect the asymptotic convergence rate and they
automatically turn themselves off as the grid is refined.

We use
a recovery version of DG for diffusion terms in the collision operator
and demonstrated some attractive properties of that
algorithm\cite{Hakim:2014}.
While this paper focuses on the electrostatic limit, we have also done
linear tests of electromagnetics, and discovered that in
order to handle magnetic fluctuations efficiently, it was important to
use consistent spaces, so that the basis functions for
$\grad_{||} \phi$ and $A_{||}$ are in the same space and the numerical
representation is able to allow (but not require) $E_{||} = 0$ at all
scales.


As we develop this code, we have undertaken various
test problems and lower-dimen\-sion\-al problems.  We did a series of 2D
tests of the generic energy-conserving properties of the algorithm for
Hamiltonian problems, and 2D tests of the parallel dynamics and
perpendicular $E \times B$ nonlinearity of gyrokinetics.  We did a
gyrokinetic simulation of a 1D test problem involving propagation of a
heat pulse along the SOL (with parameters chosen to model an ELM on
JET), and found good agreement with previous PIC and Vlasov
codes\cite{Shi:2015}.  The 1D code is orders of magnitude faster than
a full PIC code because gyrokinetics (using a sheath model) does not
have to resolve the Debye length or plasma frequencies.
%
One of the first 3x/2v test cases was in a thin flux tube in a
simplified toroidal
geometry with bad curvature, verifying that the linear growth rate for ETG
modes can be properly reproduced and demonstrating nonlinear saturation
of the turbulence.  Though \gke\ is a non-local full-F
code, it can run with a thin radial domain and periodic boundary
conditions for benchmarking with core local gyrokinetic codes.
Surprisingly, we found that periodicity must be applied at fixed
$v_\Perp$ and not $\mu$ even in the limit of a very thin domain or there
are significant errors in the free energy balance.
One might think that a simpler useful system for studying ITG turbulence
driven by bad-curvature would be a local 2D limit (ignoring the parallel
dynamics).  However, careful energy analysis has shown that this
requires direct dissipation (such as perpendicular
viscosity) to act on the energy in the electrostatic potential, or else
there will be a secular increase in the $E \times B$ kinetic energy,
with a rate of increase that is proportional to the turbulent heat flux.

\ignore{

\begin{figure}
  \setkeys{Gin}{width=1.0\linewidth,keepaspectratio}
  \vspace{-0.25in}
  \incfig[width=0.8\textwidth]{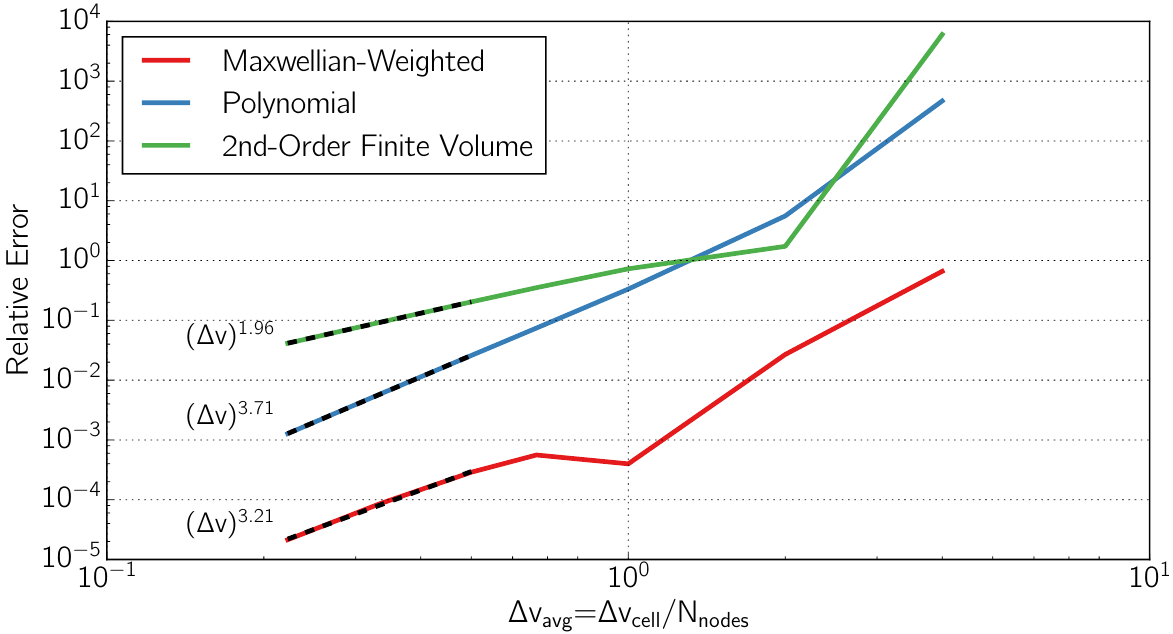}
  \caption{Relative error vs.\ velocity resolution for a Spitzer-H\"arm
    parallel heat flux problem, for 3 different algorithms: a standard
    2cd-order finite volume method, DG with standard piecewise-linear
    basis functions, and DG with exponential-weighted piecewise-linear
    basis functions.}\label{Maxwellian-convergence}
\end{figure}

} 

\vspace{-0.5em}
\subsection{Exponentially-weighted DG Basis Function}

\begin{wrapfigure}[18]{R}{0.6\textwidth}
\centering 
  \vspace{-1.em}
\includegraphics[width=0.6\textwidth]{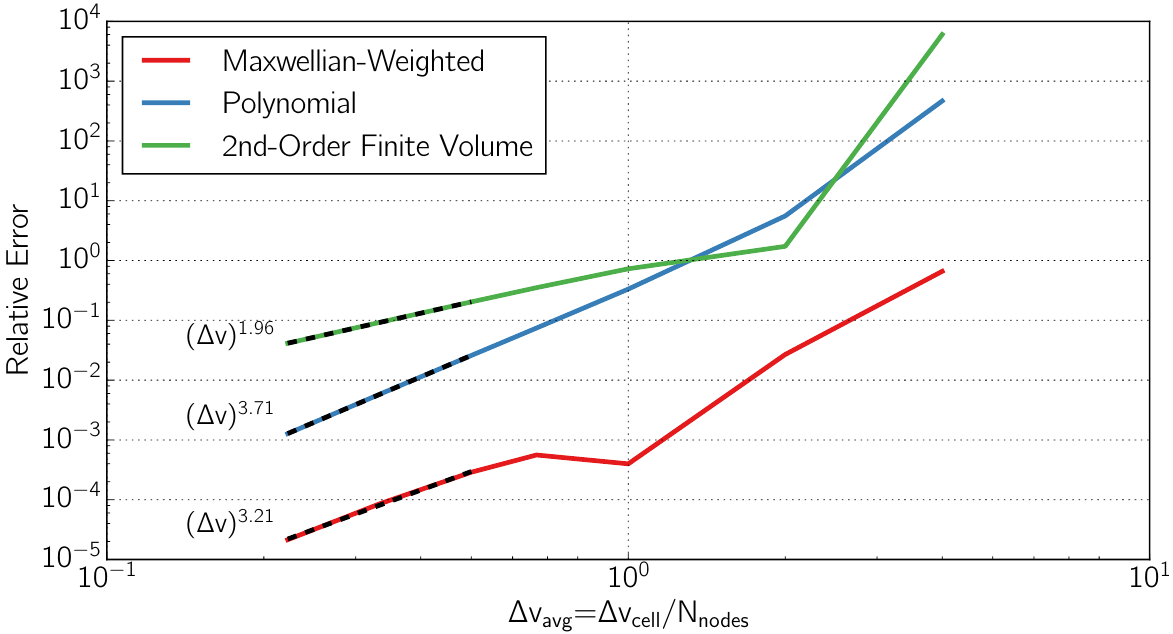}\hfill
\caption{Relative error vs.\ velocity resolution for a Spitzer-H\"arm
    parallel heat flux problem, for 3 different algorithms: a standard
    2cd-order finite volume method, DG with standard piecewise-linear
    basis functions, and DG with exponential-weighted piecewise-linear
    basis functions.}\label{Maxwellian-convergence}
\end{wrapfigure}
%
%
%
%
%
We developed a novel version of DG that uses exponentially-weighted (or
Maxwellian-weight\-ed) basis functions while still preserving exact
particle and energy conservation.  A key to preserving the conservation
properties is the choice of weight in the inner product used for the
error norm.  Consider a general equation of the form $\partial f(v,t) / 
\partial t = G[f]$, where standard DG expands $f = \sum_k f_k(t) b_k(v)$ in each cell
and chooses $\dot{f}_k = df_k/dt$ to minimize the error $\epsilon^2 =
\int dv (\sum_k \dot{f}_k b_k - G)^2$.  While the standard conservation
properties hold if polynomial basis functions are used, they are lost if
the basis functions are exponential weighted, $b_k(v) = W(v)
\hat{b}_k(v)$, where $W(v) = \exp(-\beta v^2)$ and
$\hat{b}_k$ are polynomials in $v$.  Choosing $\dot{f}_k$ to minimize a
weighted error defined as $\epsilon^2 = \int dv (1/W(v)) (\sum_k \dot{f}_k b_k
- G)^2$ will then recover the standard conservation laws.
Generalizations of this can be used to ensure positivity of $f(v,t)$.
Basis functions capable of varying
exponentially fast can represent certain features in the solution more
efficiently, as demonstrated in Fig.~\ref{Maxwellian-convergence} for a
1D test case involving parallel heat conduction, an important problem in
the SOL.  The Maxwellian-weighted DG method achieves the same error of
$10^{-2}$ with 4 times fewer grid points in this 1D problem, which is 4
times less computational cost (assuming memory bandwidth is the
limiting factor), or 16 times faster in the 2 velocity dimensions of
gyrokinetics.  This has been tested in a stand-alone code and future
work can merge this into the main code.

%


Another approach that might lead to significant improvements in
efficiency, particularly in higher dimensions, is use of sparse grid
quadrature and fewer basis functions.

\vspace{-0.5em}
\section{Formulation of model equations and boundary conditions}

We have made a number of simplifying assumptions in the equations and
the geometry for now in order to accelerate progress, but the physical
model still retains key features of edge turbulence that have been
computationally difficult in the past and are sufficient to demonstrate
the overall feasibility of this approach.  We have done simulations with
a helical magnetic geometry (i.e., a toroidal and vertical magnetic
field, such as in the Helimak and Torpex devices),
which can be used as a simple model of the
open-field line region of the scrape-off-layer in tokamaks.  
\ignore{
We have
done both open-field line simulations (with a source to model the
turbulent transfer of plasma from closed to open field lines in the SOL)
and closed field line simulations (with periodic radial boundary
conditions).  (An upgrade to handle both regions simultaneously
including a separatrix is left for future work, but the helical
open-field line model is an interesting place to start studying the
properties of turbulence in the SOL region, since it includes the
toroidal bad curvature drive and rapid parallel losses.  It can be used
to study questions of how much SOL turbulence might be expected to
broaden the power deposition width on the divertor plates or the effects
of reduced recycling by lithium walls.  While we have done helical
simulations,
}
In this paper we focus on results just with a straight
magnetic field such as in the LAPD device.

We use a gyrokinetic equation in the long wavelength limit with straight $\mathbf B$ field,
\vspace{-0.5em}
\begin{align}
\frac{\partial f}{\partial t} + \frac{\partial}{\partial z} \left(
v_{||} f \right) + \nabla \cdot \left( \vec v_E f \right)  +
\frac{\partial}{\partial v_{||}} \left( \frac{q}{m} E_{||} f \right) =
C[f] + S
\end{align}
where $f(\vec x, v_{||}, \mu, t)$ is the guiding center distribution
function, $\vec v_E$ is the $E \times B$ drift, $C[F]$ is a
Lenard-Bernstein model collision operator, and $S$ is a source that
balances losses to the end plates.
The electric potential is
determined by the long-wavelength gyrokinetic Poisson equation
$ -\grad_{\Perp} \cdot ( \epsilon_\Perp \grad_\Perp \phi) 
= \sum_s q \int d^3 v f$,
where $\epsilon_\Perp(\vec x) = c^2 / v_{A0}^2 = c^2 4 \pi n_0(\vec x)
  m_i /B^2 $
is the plasma perpendicular dielectric coefficient and is assumed to be
time-independent for now.  (This can be generalized to allow time
varying density in $\epsilon_\Perp$ if a second order contribution
$-v_E^2$ to the Hamiltonian is kept.)
\ignore{
(The gyrokinetic Poisson equation
is a statement of quasineutrality, with the RHS being the
charge density of guiding centers $\sigma_{\rm gc}$, and the LHS being
the negative of the polarization charge density $\sigma_{\rm pol}$ from
the ion polarization drift. I.e., we will always have the total charge
$\sigma_{\rm gc} + \sigma_{\rm pol} = 0$.  The polarization charge
density is related to the vorticity in fluid models of plasma
dynamics.)
} 
The RHS of the gyrokinetic Poisson equation is the guiding center charge
density $\sigma_{\rm gc}$, 
and the LHS is the negative of the polarization charge
density.

\begin{figure}[bt]
  \setkeys{Gin}{width=0.8\linewidth,keepaspectratio}
  \vspace{-0.5em}
  \incfig{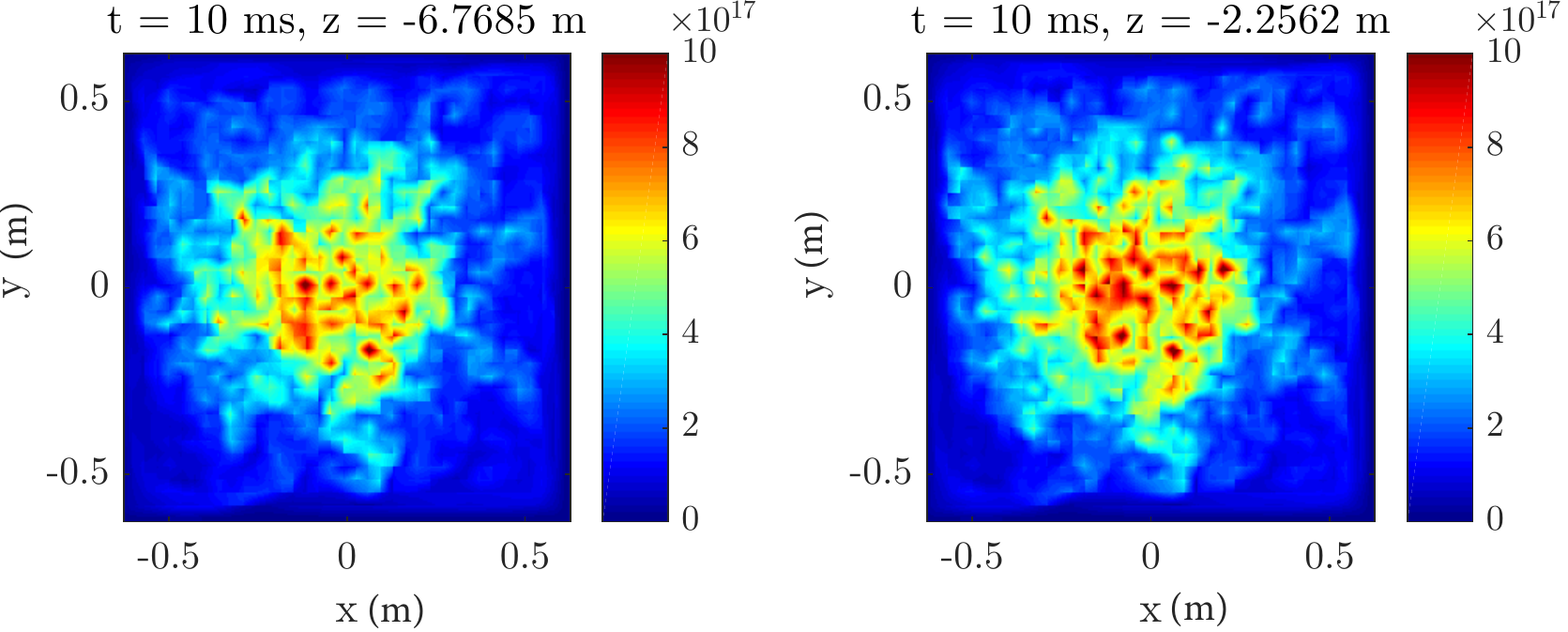}
  \caption{Contour plots of density fluctuations in the $(x,y)$ plane at
    two axial locations, from gyrokinetic turbulence simulations with
    the \gke\ code, for parameters similar to the LAPD device (UCLA).
    \gke\ uses advanced algorithms including discontinuous Galerkin
    methods methods to help with the computational challenges of edge
    turbulence.}\label{LAPD-turbulence-color}
\end{figure}

When solving the gyrokinetic Poisson equation, we use the boundary
conditions that $\phi = 0$ on the side walls, at $x=0,L_x$ and
$y=0,L_y$.
(This also means that there is no $E \times B$ loss of
guiding centers into the side walls, the only loss of guiding centers is
to the end plates at $z=0, L_z$.)  Given the guiding center charge
density $\sigma_{\rm gc}(\vec x)$, this then uniquely determines the
potential $\phi(\vec x)$ everywhere inside the plasma.  If we start with
a plasma that initially has $\sigma_{\rm gc} = 0$ so $\phi = 0$, then
electrons will be lost to the end plates faster than ions.  This leaves
behind a net positive guiding center charge and causes $\phi$ to rise.
Because the $\grad_\perp$ in the gyrokinetic Poisson equation only
involves perpendicular gradients, there will be a discontinuity between
the positive potential in the plasma and $\phi=0$ on the end plates
(assumed here to be grounded to the side walls).
This jump represents the potential jump across the sheath region, which
has a tiny width of order the Debye length, a scale 
along a field line that is not resolved in standard gyrokinetics.
Electrons without enough parallel energy will be reflected back into the
plasma (for a regular positive sheath).
I.e., defining the sheath potential at $z=0$ as $\phi_s(x,y) =
\lim_{\delta \rightarrow 0} \phi(x,y,\delta)$, we define a
cutoff velocity $v_c(x,y) = \sqrt{2 e \phi_s(x,y) / m_e}$ and impose the
boundary condition for incoming electrons at $z=0$ as
$f_{e}(x,y,0,v_{||}, \mu, t) = f_{e}(x,y,0,-v_{||}, \mu, t)$ for $0 <
v_{||} < v_c$, and $f_{e}(x,y,0,v_{||}, \mu, t) = 0$ for $v_c <
v_{||}$.  (There is no need for boundary conditions on the outgoing part
of $f$ for $v_{||}<0$.)

These boundary conditions are the gyrokinetic equivalent of the boundary
condition on electron fluid velocity used in early edge fluid
turbulence simulations (such as \cite{Rogers2010,Friedman2013}),
$V_{||e} = c_s \exp(\Lambda - e \phi_s / T_e)$, though unlike the fluid
case there is no assumption that the electrons are Maxwellian.
The sheath potential
will eventually rise until the mean electron flux is equal to the mean
ion flux to the end plates, though these boundary conditions allow the
net current at any particular location to fluctuate self-consistently,
with return currents flowing through the wall.
This differs from simple versions of a logical sheath that
force $j_{||} = 0$ everywhere on the end plate.  There are various
improvements to these boundary conditions that could
be explored in the future, such as extensions for a magnetic
pre-sheath for an inclined magnetic field 
\cite{Loizu2012, Geraldini2016},
extensions to better handle boundary layers that 
may form from polarization currents near the side walls,
and modifications if the ion acceleration from the source region
to the end plates is not large enough to exceed the Bohm sheath
criterion.  Extensions could eventually be made to treat some of the
features of the LAPD experiment in more detail.
\ignore{ such as the biasing of
various components and 
the generation of $\sim 50$ eV energetic electrons at
the hot cathode that ionizes and heats the plasma.
} 

%

\begin{figure}
  \setkeys{Gin}{width=0.48\linewidth,keepaspectratio}
  \vspace{-0.5em}
  \incfig{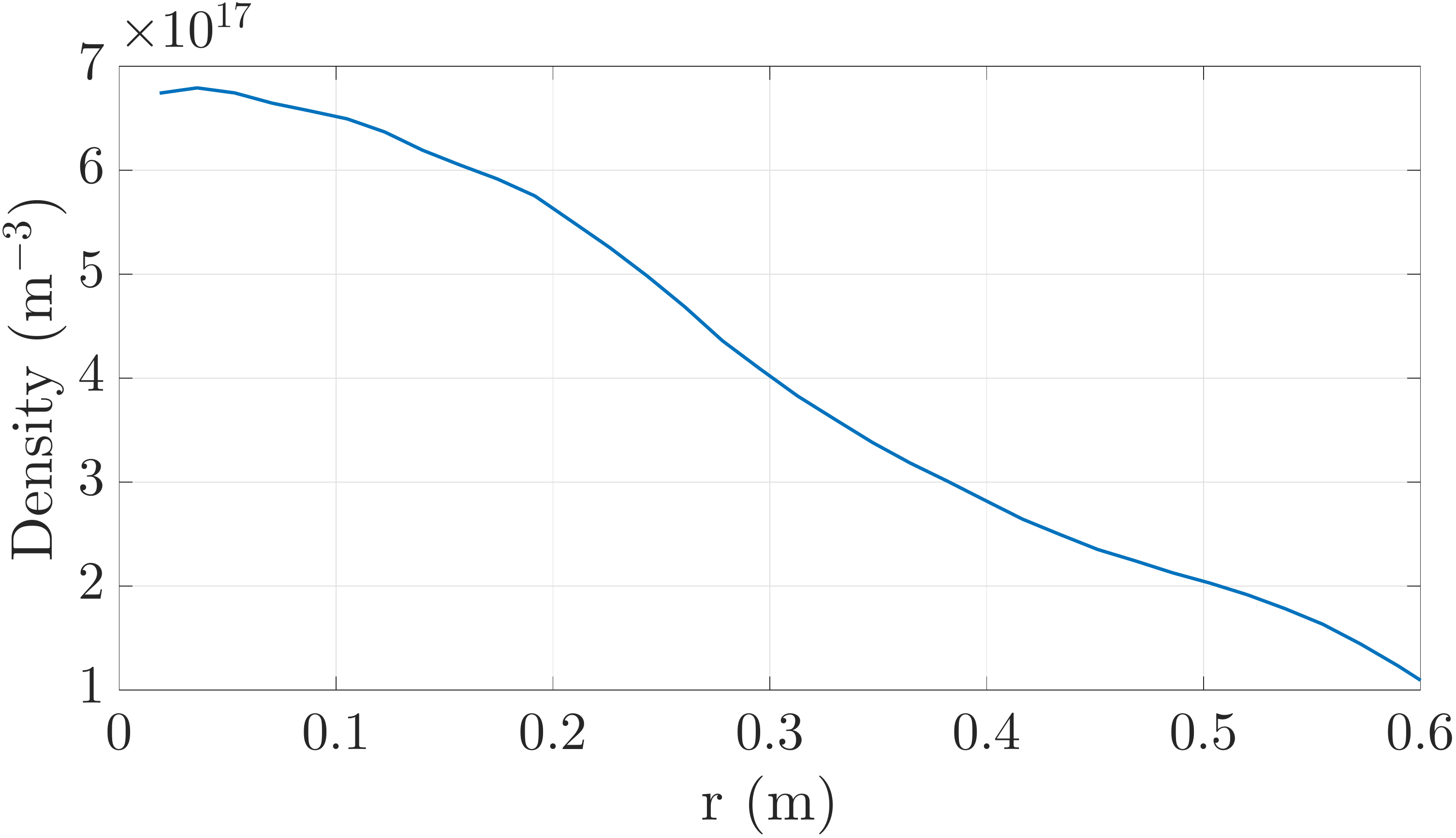}
  \incfig{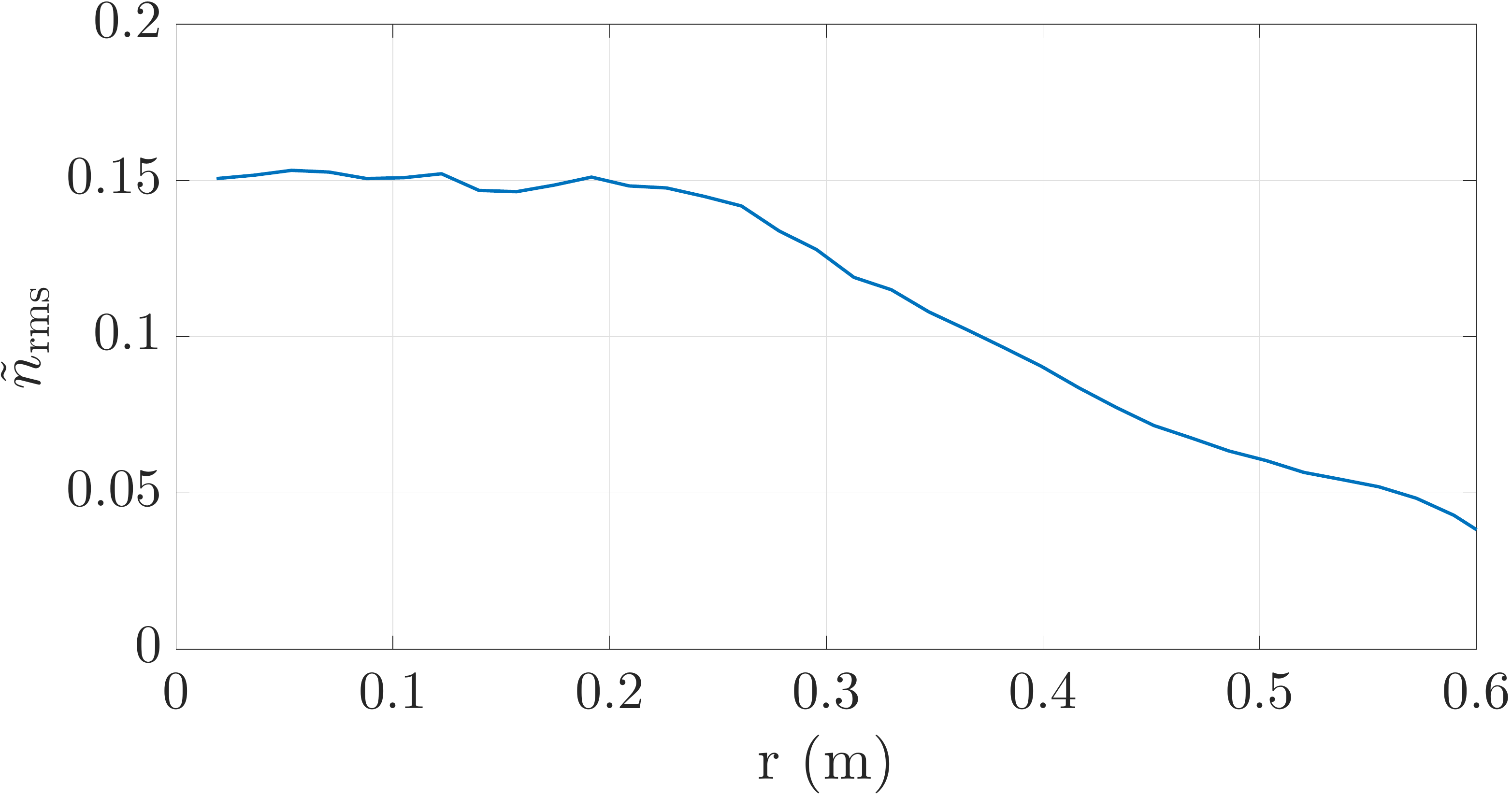}
  \caption{Radial profiles of the mean density $n_0(r)$ (left) and the
    rms density fluctuations (right, normalized to $n_0(0)$).
    The fluctuation amplitudes from these initial simulations are
    qualitatively similar to observations in LAPD.
    Future simulations will explore the impact of improved models of the
    LAPD source and setup.}\label{LAPD-den-profiles}
\end{figure}

For the above gyrokinetic equation and Poisson equation, one can show
that the total energy $W_{\rm tot} = \sum_s \int d^3 x \int d^3 v f (1/2) m v_{||}^2 
   + \int d^3 x ( \epsilon_{\Perp} / 8 \pi ) | \grad_\Perp \phi |^2 $,
which is the sum of parallel kinetic energy and perpendicular $E \times
B$ flow energy, satisfies the equation
\begin{align}
\frac{d W_{\rm tot}}{dt} = P_S - \int d S_{||} \sum_s \int d^3 v f (1/2) m
v_{||}^3 - \int d S_{||} \phi j_{||}
\end{align}
where $\int d S_{||}$  represents an integral over the surface area of the
parallel end plates, using the potential just inside the plasma at the
entrance to the unresolved sheath region.  On the RHS, the first term
$P_S$
represents the input power from the source, the second term is the
kinetic energy flux 
to the top of the sheath, and the third term represents the acceleration
of ions and the deceleration of electrons by the sheath before they hit
the wall.  If the integrated $j_{||}$ through the sheath is non-zero,
this implies the guiding center charge in the plasma is changing and
it can put energy into $E \times B$ flows.

%

\ignore{

\begin{figure}[htb]
  \setkeys{Gin}{width=1.0\linewidth,keepaspectratio}
  \vspace{-0.25in}  
  \incfig{ETG-3D-example.png}
  \caption{The electrostatic potential in the radial-binormal plane
    for a test problem of bad-curvature-driven ETG turbulence, running
    the full-F non-local \gke\ code in a periodic local limit,
    illustrating its 3x/2v capability.}\label{ETG-turbulence} 
\end{figure}

} 

\vspace{-0.5em}
\section{First Continuum Gyrokinetic Simulations of Edge Turbulence}

\vspace{-0.5em}
The LArge Plasma Device LAPD\cite{Gekelman1991,Carter2009} is a
linear plasma device that has been used to study a wide range of
plasma phenomena including turbulent radial transport.
\ignore{The vacuum vessel is 18 m long and 1 m in 
diameter.}
It uses a hot cathode and mesh at one end that creates $\sim
50$ eV electrons that ionize and heat the plasma.  Our code could
eventually do a more detailed treatment of these processes, but for now
we just use a fixed plasma source and model sheath boundary conditions
as described in the previous section.

For our initial simulations, we mostly followed the setup used in the
early fluid simulations of LAPD by Rogers and Ricci\cite{Rogers2010},
with some variations.  The typical sound speed $c_s = \sqrt{T_{e0}/m_i}
= 1.2 \times 10^{4}$ m/s and sound gyroradius $\rho_s = c_s /
\Omega_{ci} = 1.25 \times 10^{-2}$ m for $T_{e0} = 6$ ev and the domain
size was $(L_x,L_y,L_z) 
= (1.25, 1.25, 18)$ m.
\ignore{
The density fuelling rate of the source $S$ was close to a flattop
profile, with a value of $0.03 \, n_0 c_s / R_0$ for 
$r < r_s = 0.25$ m, and rapidly dropping to 0
outside of that, where $n_0 = 2 \times 10^{18} /{\rm m}^3$ and $R_0 =
0.50$ m.}
The electron and ion sources were taken to be a
Maxwellian in velocity with temperatures of close to $T_{Se} = 6$ eV and
$T_{Si} = 1$ eV in the main fuelling region.
These initial simulations were done with a resolution of $(N_x, N_y,
N_z, N_{v_{||}}, N_\mu) = 72 \times 72 \times 8 \times 12 \times 6$ node
points (piecewise linear basis functions were used with 2 nodes in each
dimension per DG cell).  More recent simulations used
a resolution of $72 \times 72 \times 20 \times 20 \times 10$ (this gives
200 velocity grid points per spatial grid point).


A snapshot of the density profiles 
at two different $z$
locations is shown in Fig.~\ref{LAPD-turbulence-color}, at $t=10$ ms
(about 13 sound times $L_z/2 c_s$).  These fluctuation features
are qualitatively similar to the fluid simulations in
Ref.\cite{Rogers2010}.  We have not yet included explicit viscosity, and
ion-ion or ion-neutral viscosity may smooth the small scales some.


\ignore{
A snapshot of the density profiles $n_e(x,y,z)$ at two different $z$
locations is shown in Fig.~\ref{LAPD-turbulence-color}, at $t=1.649
\times 10^{-3}$ s, corresponding to about 3 sound transit times using
$c_s = \sqrt{(T_e+T_i)/m_i}$.  (This particular plot is actually from
a simulation with $T_{Se} = T_{Si} = 3$ eV, but the characteristics are
similar to the case in Fig.~\ref{LAPD-den-profiles}, which is with the
$T_{Se}=6$ eV, $T_{Si}=1$ parameters above.)  These fluctuation features
are qualitatively similar to the fluid simulations in
Ref.\cite{Rogers2010}.
} 

The mean density profile and the RMS density fluctuation amplitude
profile is shown in Fig.~\ref{LAPD-den-profiles}.  The latter is qualitatively
similar to the observation of $\sim$10\% fluctuations in the LAPD
experiment reported in Fig.~2 of \cite{Friedman2013}, but more detailed
simulations for that specific experiment need to be done for careful
comparisons.
%
%

\ignore{
We have also done simulations with a toroidal magnetic field (and
associated curvature and grad $B$ drifts) of the Torpex device, which we
plan to report on in the future.  This includes the bad curvature drive
of toroidal instabilities.  The parameters and setup are similar to the
fluid simulations of Ref.\cite{Ricci2010}.  While the latest fluid
simulations\cite{Ricci2015} of Torpex can reproduce a number of key
features in the experiment, the fluid simulation turbulence level is
about a factor of 2 lower than observed.  This is an interesting area
for future research.
} 

\ignore{

We will show results from simulating turbulence in an SOL in a
simplified open-field-line geometry (similar to the helical
magnetic field geometry of the Helimak or TORPEX devices), which is
useful to testing the general feasibility of these numerical techniques
for edge turbulence.
Despite this
simple geometry, it includes magnetic curvature drive of instabilities
and sheath boundary conditions, and can be used to begin qualitative
studies of the physics that affects the width of the SOL, whether
ballooning instabilities might broaden the SOL in scaling to larger
devices, and the physics of low-recycling regimes that can be achieved
with lithium walls.

We are focusing on long-wavelength gyrokinetics and
making various other simplifications for now.  We are using a
Lenard-Bernstein collision operator, which is better than a Krook model
for plasma problems in that it has diffusion and drag terms in velocity
space like the full Landau collision operator, and so preferentially
damps small scales in velocity space.  It is a nonlinear
integro-differential operator on the distribution function and has the
same set of nonlinear conservation laws and an H-theorem as the full
collision operator, but is much simpler to implement because the drag
and diffusion coefficients are simple constants.

} 


\ignore{
\section{Adjoint methods}


Finally, Abel derived the adjoint of the gyrokinetic equation, which can
be used for various purposes, including study of the fastest-growing
instantaneous state due to non-normal modes (related to sub-critical
nonlinearly-sustained turbulence).
Adjoint methods have been useful in fluid turbulence
to study non-normal modes that can give rise to nonlinearly sustained
turbulence even when all modes are linearly stable, and they are of
growing interest for plasmas also.  Abel has derived the full adjoint
operator of the gyrokinetic equation, and has calculated the
instantaneously fastest-growing ITG/ETG state in the drift-kinetic
limit.  He has also explored the application of adjoint methods to
nonlinear gyrokinetic systems, where it might be useful for more
efficiently calculating the Jacobian of the sensitivity of turbulence to
various key parameters, which is important for design optimization and
for a stable implicit treatment of long transport-time-scale
simulations.  He is investigating methods from other fields for applying
adjoint methods to chaotic dynamical systems on a long time scale.

} 

\vspace{-0.5em}
\section{Conclusions}

We have presented first results from the \gke\ code of 5D
continuum gyrokinetic simulations on open field lines with model sheath
boundary conditions.  Results for LAPD are shown.  We are
using various simplifications at present, such as a simple helical
magnetic field model, but this is sufficient to demonstrate
the overall feasibility of this approach.
This simplified system contains some of the key physics of
the SOL region, such as the bad curvature drive of toroidal
instabilities, rapid parallel losses to the divertor plates and
interactions with sheaths, and so it can be useful to begin physics
studies about the nature of SOL turbulence, such as why doesn't this
turbulence spread power more widely on divertor plates, and what are
the effects of reduced recycling with lithium?  There is some numerical
heating in our present code from the correction step used to preserve
positivity everywhere within a DG cell, and we are working on ways to
improve this.  
There are a number of other ways that \gke\ could be
improved in the future, including exponentially-weighted basis functions
or other algorithmic improvements,
more
detailed physics including atomic physics,
and extensions to general geometry to handle open
and closed field line regions simultaneously.

\vspace{-0.7em}
\section*{Acknowledgments}
We thank P. Ricci, J. Loizu, T. A. Carter, B. Friedman, F. Jenko, B. D. Dudson,
and E. Havl{\'\i}{\v c}kov{\'a} for helpful suggestions and information,
and thank
J. Hosea, P. Efthimion, A. Bhattacharjee, M. Zarnstorff, and S. Prager
for support of this research direction.  This work was supported by the
U.S. Department of Energy through the Max-Planck/Princeton Center for
Plasma Physics, the SciDAC Center for the Study of Plasma
Microturbulence, and Laboratory Directed Research and Development
funding, at the Princeton Plasma Physics Laboratory under Contract
No. DE-AC02-09CH11466.

{
\renewcommand{\baselinestretch}{0.5}

\setlength{\parskip}{0em}

\bibliographystyle{iaea}

\vspace{-0.8em}

\small 
\bibliography{gkeyll}

}
\end{document}